\documentclass[a4paper,onecolumn,11pt]{article}
\pdfoutput=1
\usepackage[utf8]{inputenc}
\usepackage[english]{babel}
\usepackage[T1]{fontenc}
\usepackage{amsmath}
\usepackage{hyperref}

\usepackage{tikz}  
 
\usepackage{fullpage}
\usepackage{amsfonts,amssymb}
\usepackage{amsthm}
\usepackage{tikz}
\usetikzlibrary{calc}
\usetikzlibrary{arrows}
\usepackage{tikz-3dplot}
\usepackage{color}  
\usepackage{hyperref}
\usepackage{bm}
\usepackage{float} 
\usepackage{appendix} 
\usepackage{lscape}
\usepackage{cite}
\usepackage{mathrsfs}
\usepackage{appendix}
\usepackage{setspace}
\usepackage{mathtools} 
\usepackage{natbib}

\theoremstyle{definition}

\begin{document}

\title{Quantum Field Theory and the Limits of Reductionism}

\author{Emily Adlam  \thanks{Philosophy Department and Institute for Quantum Studies, Chapman University, Orange, CA92866, USA \texttt{eadlam90@gmail.com} }}

\maketitle

\begin{abstract}
I  suggest that the current situation in quantum field theory (QFT) provides some reason to question the universal validity of ontological reductionism. I argue that the renormalization group flow is reversible except at fixed points, which makes the relation between large and small distance scales quite symmetric in QFT, opening up at least the technical possibility of a non-reductionist approach to QFT.  I  suggest that some conceptual  problems   encountered within QFT may potentially be mitigated by moving to an alternative  picture in which it is no longer the case that the large supervenes on the small. Finally, I explore some specific models in which a form of non-reductionism might be implemented, and consider the prospects for future development of these models.  

\end{abstract}

There is an almost universal consensus in modern science that the most fundamental things in nature appear at the smallest scales,  so  everything  happening at large distance scales supervenes on whatever is happening at smaller distance scales. This idea has been a part of science for a very long time, at least since the work of Democritus \citep{sep-democritus}. Indeed it seems likely that ideas of this kind featured in common-sense `folk science' long before any kind of formal science existed - though as \cite{midgley2005myths} notes,  the seventeenth-century invention of the microscope significantly bolstered the credibility of such ideas.

But in this article, I will argue that our current understanding of quantum field theory (QFT) provides some reason to question the universal validity of this reductionist assumption. It is not my intention to offer a conclusive argument to the  effect that the supervenience of the large on the small definitely fails within QFT - with so much still unknown about physics at the Planck scale and below, it would be premature to say anything too definitive on this point. So my goal    is simply  to argue that the standard reductionist picture should not be taken for granted in the regimes relevant to QFT.   I will try to make the case that there are some interesting features of QFT which may point in this direction, and that  some conceptual  problems   encountered within QFT may potentially be mitigated by moving to an alternative  picture in which it is no longer the case that the large supervenes on the small. I will also try to show that a commitment to the supervenience of the large on the small is not forced on us by the structure of QFT as we presently understand it.

Now, of course one obvious way to deny that supervenience of the large on the small is to adopt some kind of operational, empiricist or anti-realist approach. For example, if one adopts an idealist picture in which science is nothing but a description of human experiences, then of course nothing at all can exist at scales inaccessible to human observers, so it cannot be the case that the large really supervenes on the small given that very small scales are  inaccessible to us. However, this is not the kind of approach I will pursue here - I am interested in considering the status of reductionism within a realist picture. For after all, the   realist surely should not hold sacred any particular aspects of the traditional picture   of physical reality; in particular, the notion of `realism' should not be tied to any particular methodology for explaining physical reality, such as reductive explanations of the large in terms of the small, and therefore realists should be prepared to relax their reductionism if the evidence points in that direction.

\section{Terminological Notes}

Before going on, I will introduce some terminology. First, note that in this article I will use the word `reductionism' to refer specifically to a view sometimes known as `microphysicalism' \citep{Huttemann2003-HUTWWW-3} or   `petty reductionism' \citep{Weinberg+2001+107+122}, referring to the idea that what happens at large distances is entirely determined by the behaviour of the constituents of nature at the smallest possible scales. That is, I willl be specifically concerned with reductionism as it pertains to \emph{physical size}. The term `reductionism' is sometimes used to refer to more general positions, such as the idea that there exists some physical supervenience basis for all of reality, or the belief that all mental events are reducible to physical events \citep{Velmans1998-VELGTR}, but I am not addressing these views here - the failure of reductionism in the sense in which I will use that term would not entail that there is no universal supervenience basis or that mental events are not reducible to physical events. 

Second, in this article I will distinguish between  `methodological reductionism' \citep{FRANKLIN2023123}, which refers to the idea that   seeking to derive macroscopic dynamics from microscopic dynamics together with suitable auxiliary assumptions is the best \emph{method} for studying a certain kind of system, and `ontological reductionism,' which refers to the ontological claim that all goings-on at large scales fully supervene on goings-on at smaller scales, regardless of whether or not that reduction can be written down by  human scientists. Philosophical presentations of anti-reductionism have often focused on arguing against methodological reductionism, as in the work of  \cite{BATTERMAN2023130} using examples from many-body physics to show that `bottom-up' derivations are not always the most methodologically fruitful approach, or the work of \cite{cd3bcfc7-c1f3-3ac4-a548-136b57165545} on effective field theories, or various arguments that have been made opposing  the idea that the best explanation is always the one at the smallest possible scales \citep{8200f7e6-d2b7-3241-845a-13760a84c7a0,doi:10.1093/bjps/axy015}. However, these issues pertaining to methodology and explanation are in part a function of pragmatic human interests, whereas in this article I will be addressing the issue of \emph{ontological} reductionism, so I am defending the stronger claim that as a matter of physical fact, independent of any pragmatic or explanatory considerations, it might not always be the case that in the regime of QFT large-distance effects entirely supervene on small-distance effects. 
 
Third, a note about the term `fundamentality.' It is easy to get the feeling in debates about `fundamentality' that there is really no objective fact about what is  fundamental, and indeed that this is not a very useful question in any case. So it is important   to be clear about what we consider the significance of `fundamentality'  to be. Specifically, in this article I will understand questions about fundamentality  as questions about whether a complete physical description of happenings at one scale can be expected to supervene entirely on a complete physical description of happenings at some other scale, and  the reason I consider these questions important is  because they are relevant to decisions about where to go looking for new physics. For although ontological and methodological reductionism  are distinct, nonetheless there is clearly a connection between them - in particular, if no ontological reduction is possible in some regime, then it is likely that the reductionist methodology will not be a good way of studying that regime.  For example, if we believe that ontological reductionism is true in the regimes relevant to QFT, then it seems natural to think that a complete, consistent theory of the physics at the smallest distances must exist, since large-distance physics could not supervene on it if it did not exist; so this point of view provides strong motivation either to formulate a consistent, non-perturbative continuum QFT, or to seek some alternative account of small-distance physics. Whereas if we believe that ontological reductionism is false in the regimes relevant to QFT, then the physics of the smallest scales can potentially be regarded as simply an unphysical limit that will never actually be reached,  so  it need not necessarily be the case that there exists any complete, consistent description of the smallest-scale physics, and physicists should instead be focusing on understanding how small-scale behaviour can arise from some larger-scale constraints. 

With that said,   the relationship between ontological and methodological reduction is not straightforward. It is true that if ontological reductionism is false in QFT  it is likely that methodological reductionism will not be a good strategy for making further scientific progress in this regime, but also if ontological reductionism is true in QFT then it could \emph{still} be the case that methodological reductionism is not a good strategy for making progress in QFT. After all, in general scientists will be able to write down an exact reduction from large scale dynamics to small scale dynamics only if there exist some stable, relatively autonomous large-distance dynamics, and those dynamics can be derived from some relatively simple small-distance dynamics plus relatively straightforward auxiliary assumptions about things like initial conditions, parameter values, and/or Nagelian bridge laws. Yet even when there is perfect supervenience in an ontological sense, it is entirely possible that one or more of these things  could fail to hold -   for example,   the auxiliary assumptions needed to derive macroscopic dynamics from microscopic dynamics could simply be too complex to be formulated by human scientists. Thus simply knowing whether  ontological reductionism holds in some regime does not immediately tell us what methodology we should pursue to study that regime. However, it does seem true to say that if we were to learn that ontological reductionism is definitely \emph{false} in the regimes relevant to QFT, that would have a significant impact on the way we would thereafter study those regimes. So I emphasize that it is not my intention in this paper to engage in a  semantic debate about the meaning of the word `fundamental' - I am interested in a concrete scientific question about the appropriate methodology for  future physics research.

Finally, a technical note: in quantum mechanics, energy  behaves like the reciprocal of distance. This can be understood in terms of the de Broglie relation, which specifies that the wavelength $\lambda$ of a probe system is given by $\lambda = \frac{\hbar}{p} $ where $p$ is the momentum of the probe. No probe can give us information about phenomena smaller than its own wavelength,  so  as we go to smaller scales we need probes of larger momenta or equivalently larger energies to probe phenomena at those scales. Thus in QFT `small distance' and `high energy' are often used somewhat interchangeably, and likewise for `large distance' and `low energy.' In this paper I will follow this convention, although when possible I will emphasize the description in terms of distance scales, since I am particularly concerned with reductionism as it pertains to physical size.

\section{Background}

In this section I will give some background on quantum field theory and renormalization. Readers who are already familiar with these concepts may wish to go straight to section \ref{intro}.

Quantum field theory is, in principle, simply the result of applying quantum mechanics to fields such as the electromagnetic field. In this approach a field is modelled as being composed of an individual quantum system at every point of spacetime, each having some number of degrees of freedom: for example, a `scalar field theory' has one degree of freedom, i.e. a scalar, at every point. A field theory is characterised by a Lagrangian,  $L$ which typically includes a kinetic term  for each field, as well as terms representing the interaction of a field with itself or with other fields, each of which  is multiplied by a `coupling coefficient' indicating the strength of the interaction. 

  Usually we extract empirical predictions from a quantum field theory by calculating transition amplitudes (giving the probability for a certain input state of the field to transition into a certain output state) or correlation functions (giving the correlation between the field values at two spacetime points). In principle this is done using a path integral, where we integrate over all possible paths between the points weighted by $\exp^{iS}$, with the action $S$ for a path given by integrating the Lagrangian over the path. But unfortunately,   the only correlation functions of this kind that we can calculate explicitly are those for field theories where the Lagrangian is at most quadratic in the field - for example, a field theory without interactions. For interacting field theories (i.e. all theories representing fields that occur in our actual world) we don't know how to do this calculation exactly, so instead we typically make predictions using a perturbative expansion, in which we expand around the more easily calculated free theory. This expansion is represented pictorially by Feynman diagrams, with diagrams having more loops representing higher-order terms in the perturbative expansion. 

However, this method has a serious flaw:  in many cases the  contribution to the perturbative expansion associated with each term turns out to be infinitely large, because we are integrating over arbitrarily low distances or correspondingly high energies. But this can't be right, since correlation functions correspond to actual physical observables and thus must be finite. This indicates that we are going to have to introduce some kind of cutoff in distance or energy, but how can we do that in a physically realistic, non-arbitrary way? The technical process by which this is achieved is known as renormalization.

\subsection{Renormalization \label{renormalization}} 

 I will distinguish two different but related approaches to carrying out  renormalization, following the discussion of these approaches in   \cite{RIVAT201923} (see \cite{peskin,Collins_1984,schwartz2014quantum,collins2009problem} for other ways of implementing these approaches). 
 
 I begin with the continuum approach, which proceeds in two steps,   First, we consider the value of some physical quantity $\Lambda$ (e.g. a correlation coefficient or `cross-section'), and we assume that it is a function of a set of Lagrangian parameters $\alpha$ (coupling coefficients and/or masses),  as well as the energy scale $Q$ at which the experiment is performed. The usual perturbation theory calculation for this quantity would involve integrating over all of space, or equivalently over all possible internal energies, but this   would give an infinite value, so  we regularise the theory by integrating only up to an arbitrary cutoff energy $E$ (or equivalently, down to some minimum distance $d$).   In order to ensure that physical quantities have the right dimension, we must introduce the cutoff together with an additional variable $\eta$ with the same dimensions as $E$, so $E$ only appears in expressions of the form $\frac{E^2}{\eta^2}$. Now, in the continuum approach we don't assume that the cutoff has physical meaning, so we need to take a limit as  $E$ goes to infinity. To determine how the parameters  $\alpha$ vary as the cutoff is taken to zero, we impose a boundary condition  which requires that  $\Lambda$  must be the same for every value of the cutoff - for after all, in the continuum approach the cutoff is an arbitrary value which is only used to regularise the theory and which doesn't correspond to anything in reality, so the predictions of the theory should not depend on its value. This requirement can be used to define a  transformation $T_1$ specifying how the parameters  $\alpha$ vary with the cutoff.  Taking the cutoff to zero, we arrive at a `bare theory' complete with precisely specifiable (albeit perhaps infinite) values for the parameters  $\alpha$. Note that, as reinforced by  \cite{pittphilsci14591}, the transformation $T_1$ does not correspond to anything empirically observable - the cutoff is just an arbitrary parameter with no physical meaning, so we can't actually test out the theory at different values of the cutoff. Therefore the boundary condition used to fix the transformation $T_1$ is not a substantive physical assertion; it can simply be regarded as  a choice of convention which happens to be convenient to define the continuum limit. 

Now, the value of  $\Lambda$  calculated from the renormalised theory, and also the renormalised parameters  $\alpha$,  will still depend on the  new variable $\eta$. In fact, since  $\Lambda$ is dimensionless, it depends only on the ratio $\frac{Q^2}{\eta^2}$, as well as the renormalised parameters - note that $E$ is necessarily replaced in this ratio with $Q$, the energy scale at which the experiment is performed, since $E$ has been taken to zero. But   $\Lambda$ is a physical observable and therefore shouldn't depend on an arbitrary choice of  $\eta$. This can be addressed by stipulating that the dependence of the correlation coefficient on $\eta$ is cancelled by a corresponding dependence of the parameters  $\alpha$ on $\eta$; and we can achieve this by setting $\eta = Q$, so by definition $\frac{Q^2}{\eta^2} = 1$. Thus the correlation coefficient now depends only on the parameters $\alpha$, while the parameters $\alpha$ will now depend on $Q$. That is, the parameters $\alpha$ are now a function of the physical energy scale, and the correlation coefficient depends on the physical energy scale only via its dependence on  $\alpha$.  So we now have a second kind of transformation, $T_2$: we can use the functions we have obtained to establish how the Lagrangian parameters $\alpha$ vary as we change the scale $Q$. 

Thus, as a side-effect of carrying out the renormalization procedure, we have discovered the scale-dependence of the Lagrangian parameters. The idea is that the observed values of the mass and the coupling constant vary according to the scale at which they are probed, because `particles' (i.e. excitations in fields) polarize the quantum vacuum around them and `\emph{the division of an excitation into ‘particle’ and ‘bits of polarization around a particle is somewhat arbitrary and, on any natural choice of convention, ends up being scale-dependent}' \citep{pittphilsci20537}. So the process of renormalization amounts to  rewriting the correlation functions in terms of the \emph{measured} masses and coupling constants, in order that we can use values obtained in experiments to actually perform calculations.
 
The second approach to renormalization, sometimes known as the effective approach, can be understood as discarding the first part of this process and focusing on the second part. The insight here is that we do not, in fact, have to take a continuum limit in order to see how the parameters should vary with scale: we can simply integrate over high-energy degrees of freedom and thus directly obtain equations for the scale-dependence of the Lagrangian parameters (i.e. obtaining a transformation similar to $T_2$, although the specific form of the equations obtained by the effective method might not always be the same as those obtained by the continuum method).  Then the renormalization transformation can be understood as  moving up through a series of what we call `effective field theories' \citep{petrov2015effective} i.e. descriptions of the system which are accurate at a certain scale but which disregard degrees of freedom at smaller distances. The transformation effectively packages all the dependence on the cutoff and the small-distance phenomena into the Lagrangian coefficients, allowing us to perform calculations for the relatively low-energy regimes that we can access even though we don't know the details of the higher-energy theory. 
 
There are several ways of interpreting the effective approach. One possibility is to think of it as encoding a tower of `effective field theories' defined at different scales, with no fundamental theory at the bottom \citep{cd3bcfc7-c1f3-3ac4-a548-136b57165545}. For an operationalist,  this bottomless tower of field theories seems like the most natural interpretation: an effective field theory is just a tool to predict the results of our experiments, so as we go to higher energies in our experiments, we move to levels of the tower defined at higher energies. The operationalist may be quite happy to say that the tower has no bottom -  if there were no technological obstacles, we could go on forever probing higher and higher energies.  But many scientific realists find the bottomless tower  less appealing, because many realists are also to some degree foundationalists, i.e. they believe that there is some fundamental level of reality on which everything else in the physical world supervenes, and from this point of view it is hard to understand what it would mean for the world to be represented faithfully by a bottomless tower of effective field theories. Although it should not be taken for granted that realists are committed to foundationalism, nonetheless many scientific realists do espouse it and thus prefer to adopt an alternative interpretation of  the effective approach which involves postulating that there is in fact a real, physical cutoff at some minimum length scale \citep{wallace2001defence,cao2019conceptual}, perhaps because QFT itself breaks down when we reach a regime where quantum gravity effects become significant (i.e. at the Planck scale). QFT does not apply below this cutoff, so we don't have to worry about whether or not the continuum theory is well-defined, but the effective field theories defined at distances larger than the cutoff are still accurate descriptions of the physics at those distances.\footnote{We note that language distinguishing approaches to renormalization varies across the physics and philosophy community. In particular, the distinction between the continuum and effective approaches we have described here is somewhat similar to the distinction made by Fraser between`QFT-without-cutoffs' (similar to the continuum approach) and `QFT-with-cutoffs' (similar to the effective approach, in the case where one assumes a physical cutoff)\cite{FRASER2011126}.}

\subsection{Reversibility   \label{intro}}

The literature is not always very clear about whether or not  the flow induced by the renormalization group transformation is reversible. Strictly speaking the term `group' should only be applied in the case of a reversible transformation, but in fact it is widely agreed that really the renormalization group should really be referred to as a semi-group, and the transformation associated with a semi-group need not be reversible. And indeed, \cite{cao2004conceptual} suggests that the renormalization group transformation is definitely irreversible at least in the condensed matter case: `\emph{Although it is easy to do the averaging (in a condensed matter system), which in general results in a different model in each step of averaging, no inverse operation would be physically possible. That is, the procedure of Kadanoff transformations is not invertible.}' Meanwhile, \cite{RIVAT201923} suggests that it is reversible in the continuum approach but not the effective approach: `\emph{The effective RG transformation obtained by decreasing $\Lambda$ is irreversible since it eliminates high energy degrees of freedom, while the continuum RG transformation obtained by varying $\eta$ is reversible since it merely amounts subtracting or adding some finite quantity in the action (i.e., to imposing a different renormalization condition).}' And \cite{ButterfieldBouatta} suggest that the reversibility is up to the theorist: `\emph{This point is not meant to suggest there is a problem, or even a puzzle, about such irreversible flows. The exact definition of the flow is up to the theorist, and of course varies from theory to theory, and according to the theorist’s purposes.}'

One source of confusion here is that there are two different things one might have in mind when asking  about features of the renormalization group transformations. On the one hand, at least from a realist point of the view it seems natural to think that there is an `actual' transformation between scales. Certainly if one is a (petty) reductionist, then one is committed to the view that there is some fact of the matter about how physics at other scales is generated from physics at the smallest possible scale; and more generally, if one finds it plausible that there is some `fundamental' scale such that the physics at other scales follows from the physics at that scale, then again there must be some fact of the matter about the specific form of the transformations by which the physics at other scales is generated from physics at the fundamental scale. Indeed, even if one is not committed to the view that some particular scale is fundamental, nonetheless  there is presumably a well-defined fact of the matter about what   physics is actually instantiated at different scales in our world, and therefore there will also be a fact of the matter about the transformation between scales, in the sense that given any specification of a standard for simplicity, we must be able to write down the simplest possible formulation of the scale transformation (which may or may not be unique). So it seems meaningful to ask whether the actual scale transformations are reversible, or at least to ask whether the actual transformations, relative to some natural standard of simplicity, are reversible.

On the other hand, in practice we use various different methods  for generating scaling transformations, and these methods are usually approximations in one way or another, meaning that they actually generate slightly different scaling transformations. So   our existing methodology does not deliver us a unique way of transforming between scales - and in particular, as noted by \cite{ButterfieldBouatta}, some methods of approximation may deliver reversible transformations while others may deliver irreversible transformations. Thus  if the term `renormalization group transformations' is understood to refer to the collection of methods rather than the real underlying relationships between scales, there is simply no fact of the matter about whether or not the transformations are  reversible. 

From the point of view of ontological reductionism, what matters is whether or not the actual underlying transformations are reversible.  But since we can only study the transformations generated by our existing methodology, we may not be able to determine directly whether or not the actual underlying transformations are reversible. Nonetheless, since the empirical evidence shows an impressive level of agreement between extant renormalization group methods for generating scale transformations and  actual observations, we have at least some reason to think that we can arrive at some conclusions about the reversibility of the `actual' transformations based on what we know about the approximate ones. 
 
With this  in mind, consider  two arguments commonly used to argue that `the' renormalization transformation must be irreversible. Argument A points out that  in the process of renormalization many different fine-grained Lagrangians can be taken to the same coarse-grained Lagrangian, and of course many-to-one transformations are always irreversible. Thus if we consider that the occurrence of this effect in flows generated by extant renormalization group methods counts as evidence that similar effects may arise in the actual underlying scale transformations, then Argument A suggests the actual  transformations should be irreversible.   Argument  B, similar to the argument made by \cite{cao2004conceptual},  points out that in the effective approach  the renormalization transformation is calculated by integrating out high energy degrees of freedom, or `coarse-graining,' and of course coarse-graining is irreversible: information is lost in this process and thus we cannot simply transform back. Thus, insofar as the actual scale transformations should also be thought of as a form of coarse-graining, it appears that the actual transformations may be irreversible. Argument B perhaps only supports the view that the actual scale transformations are   irreversible if we take it that the effective approach is the correct way to interpret the actual scale transformations; but since many-to-one transformations can occur in the continuum approach as well as the effective approach, Argument A seems to offer some reason to think that the actual transformations  are irreversible, regardless of whether the effective or continuum approach is the correct interpretation of them.

 But contrary to the conclusions suggested by Argument A and Argument B, in practice  renormalization group methods \emph{are} very often used in reverse. For example, the reverse group flow is used to calculate the way the strong coupling $\alpha_S$ of QCD will change as we go to smaller scales: `\emph{While QCD does not predict the absolute size of (the strong coupling parameter) its energy dependence is determined precisely. If (the parameter) is measured at a given scale, QCD definitely predicts its size at any other energy scale through the renormalization group equation}' \citep{Bethke_2007}. Moreover, these applications are not merely theoretical - \cite{Bethke_2007} concludes that experiments have shown `\emph{an unambiguous verification of the running of (the coupling constant) and of asymptotic freedom, in excellent agreement with the predictions of QCD.}' So it appears that in at least some practical situations, the renormalization transformations are in fact reversible and this can be empirically verified, which intuitively seems to offer  some reason to think that the actual scale transformations are  reversible. 

But how is the empirical success of the reversed renormalization group flow to be squared with Argument A and Argument B? Well, first we need to be more specific about what is meant by reversibility: for in fact, many methods of generating a renormalization group flows are  reversible  \emph{except at fixed points}. This can be seen in typical diagrams of the renormalization group flow, where we can usually move in a straightforward way in either direction along the curves defined by the renormalization group in parameter space. Only when the curves cross will the flow be many-to-one and hence irreversible, and this occurs only at fixed points, which are comparatively rare. In the words of Fisher: `\emph{we can, as it were, run backwards in the semi-group, provided there is no crossing of trajectories. And the only crossing of trajectories occurs at a fixed point, not in the vicinity. So in fact we can go backwards}'\cite{cao2004conceptual}.

This makes it clear why many methods of generating scale transformations lead to a renormalization group which is only a semi-group, even though the transformations are  mostly reversible: a true group would be reversible everywhere, but the existence of fixed points entails that the renormalization transformations will be irreversible in a few specific cases. This observation takes care of Argument A: if we suppose  that the structure of the  transformations generated by extant renormalization group methods are potentially reflective of the structure of the actual scale transformations, then it is likely to be the case that these transformations are  one-to-one except possibly at fixed points.  Thus as long as we are not starting at a fixed point, such transformations  should make it possible  to calculate a reversed flow such that small-distance physics can be determined by large-distance physics just as well as vice versa.  

But what about Argument B? After all, if the process of integrating out degrees of freedom leads to an irreversible transformation, one would naturally expect this to apply in all parts of the renormalization group flow, not only at fixed points, since every time we zoom out we  lose some information about the fine-grained state of the system. As a first response, one might posit, as does \cite{RIVAT201923}, that the renormalization group flow is reversible (except at fixed points) in the continuum approach, but not the effective approach. But this would be a surprising conclusion, because it would entail that the continuum and effective approaches are not empirically equivalent, even with respect to the evidence we currently have:  given a physical theory defined at some distance scale, the reversible continuum approach would make a specific prediction for the values of the parameters at every smaller distance scale, whereas the effective approach would  not be able to make such a prediction. And since the scale dependence of various Lagrangian parameters has been empirically observed and largely seems to match predictions, this would entail that the continuum approach is able to make a novel, verified prediction which the effective approach fails to make.

But a matter of fact,  the running of the coupling constants (towards both small and large distances) can be predicted within the effective approach too, so the successful prediction of the running of the coupling constants does not favour the continuum approach over the effective one. Yet how can this be, if the effective approach loses information by integrating out degrees of freedom? To answer this question,  we must distinguish  between the effect of coarse-graining on \emph{states} and the effect on \emph{dynamics}. Of course it is true that averaging over degrees of freedom loses information about the specific microscopic state of the system; but the \emph{dynamics} of a system are not a function of the state of the system, or indeed a property of an individual system at all, since two systems with different instantaneous states can have the same dynamics. So losing information about the microscopic state of a system does not necessarily entail that we  have lost  information about the \emph{dynamics} of that system:   the map from Lagrangians to Lagrangians induced by coarse-graining can be one-to-one even if the map from states to states is many-to-one. As Fisher puts its: `\emph{we put in as many parameters as we're interested in, field parameters, parameters that control the other ones. It's that information that we want, and that is what is retained ... It looks as though we're losing. It looks as though we're integrating out. But for the information we want for experiment, that's not the situation}' \cite{fishercao2004conceptual}.

Admittedly, it seems obvious that if we are given some coarse-grained dynamics and we are allowed to invent any exotic fine-grained dynamics we like, we will be able to come up with several different fine-grained dynamics which average out to the same coarse-grained dynamics - we can simply design these fine-grained dynamics such that the parts which differ always exactly cancel out during averaging. But in a renormalization group transformation we are not allowed to just  invent any dynamics we like: the point is that we are keeping the form of the theory the same and just changing the Lagrangian parameters. And in fact under this specific restriction, many of our methods for generating scale transformations have the consequence that everywhere but at fixed points, for a given coarse-grained Lagrangian $L_c$  there is only one fine-grained Lagrangian (with the same terms but different parameter values) which can give rise to $L_c$ under the coarse-graining transformation. For example, in many cases a coupling parameter can be shown to transform as $\lambda \rightarrow (\frac{x}{x'})^D \lambda$ where  $x'$ is the original distance scale, $x$ is the new distance scale, and the bare coupling constant has units of length$^{D}$ \citep{ButterfieldBouatta} - and clearly this transformation can be employed to move either from larger distances to smaller distances or vice versa,  so the transformation is reversible despite the fact that it was obtained by integrating out degrees of freedom. 

This  feature of the extant renormalization group transformations is striking and surely quite remarkable - it is certainly not an expected or generic feature of physics that the dynamics at a larger distance should be able to fully determine the dynamics at a smaller distance. So it seems natural to infer that the appearance of this behaviour in the   transformations generated by our extant methods may be reflective of  features  of the actual scale transformations, which suggests that the actual transformations may also be reversible at the level of the dynamics despite the fact that the coarse-graining loses information about states. In addition, using this observation, the proponent of the effective approach can make sense of the success of the continuum approach without necessarily believing in a physically realised continuum theory. For  the continuum construction can be understood as simply a convenient way of finding \emph{some} fine-grained Lagrangian which will give rise to a certain coarse-grained Lagrangian under the coarse-graining transformation, and then we can use the fact that there is usually \emph{only one} such Lagrangian to identify it with the one found using the continuum approach.  

Now, it should be reinforced that even if the actual underlying scale transformations are reversible, this   does not mean that physicists can just run a renormalization transformation backwards and immediately learn everything about physics at very small scales. For a start, some of the approximation methods available to us are not reversible. And furthermore, once we understand that  renormalization group transformations are typically reversible only because the form of the Lagrangian is so restricted, we can also see that the methods available to us might fail to be reversible   when we arrive at scales where the Lagrangian undergoes some significant change in form. For example, sometimes there is a term in the small-distance Lagrangian which does not appear in the large-distance Lagrangian, because its coupling coefficient at large distances is so close to zero that the effects of the term can be ignored. But typically the coefficient will be transformed away from zero  by the  renormalization group flow towards small distances. Thus the reversed transformation will fail to correctly predict the small-distance  Lagrangian unless we remember to include the extra term in the large-distance Lagrangian. Moreover, since the coupling coefficients of such terms are close to but not exactly equal to zero at the distances which we are able to probe, in order to implement a strictly correct reversal we would need to know the actual values of these coefficients: in practice we usually start from a large-distance Lagrangian in in which the coefficients for these terms are set to exactly zero, but this starting point cannot be exactly correct, and thus the small-distance Lagrangians we arrive at by this method will not be exactly correct either\footnote{Thanks to a anonymous reviewer for raising this point.}. To make matters worse, the number of possible terms of this kind is sometimes infinite, so we cannot realistically account for all of them. Furthermore, the renormalization transformation acts on a set of fields, and there is  no guarantee that we have included all the right fields in our large-distance Lagrangians - for example, there might be a massive particle which is important at small distances but whose corresponding field does not appear in the large-distance Lagrangian because its effects are negligible at larger distances. And of course, if there is a field missing from the large-distance Lagrangian, simply transforming the parameters in that Lagrangian via the renormalization transformations will not recover that field as we go to smaller distances. Thus it should be reinforced that there are significant practical barriers to actually `zooming in' all the way to the final continuum theory, even if the transformation is in principle strictly reversible, so the procedure of running the transformation backwards to get small-distance effects is not guaranteed to work.

\section{The Large and the Small \label{lats}}

Why do we believe that the large supervenes on the small? This is not a piece of a priori knowledge: one can at least imagine worlds in which the large does not supervene on the small  (as indeed I am attempting to do in this paper!). So in fact, the widespread conviction  that the large typically supervenes on the small is  presumably based on empirical observations. Even scientifically unsophisticated observers can see that in the macroscopic world,  `big things' are typically made out of specific arrangements of `small things,' so for example a piece of cloth is made out of many threads woven together. And as science developed, the scientists of the past discovered, through observation and associated theorizing, that macroscopic phenomena could often be accounted for satisfactorily by deriving them from  the behaviour of a relatively small number of microscopic constituents of matter which obey relatively simple rules: for example, the great variety of chemical elements can all be explained as different combinations of electrons, neutrons and protons. This methodological reductionism has been so wildly successful that the supervenience of the large on the small has become elevated to a revered and almost unchallengeable scientific principle.  

But if our faith in the supervenience of the large on the small is grounded in empirical observations, then the question of how far this supervenience principle extends is also an empirical matter. Our intuitions about this issue were developed first within the macroscopic world, and then subsequently we discovered that these intuitions could be extended all the way down to the scale of atoms and below. But when we do quantum field theory we are probing regimes far smaller than atomic radii - the Planck scale is about 10,000,000 times smaller than an average atom. And if quantum mechanics has taught us anything, it is that intuitions developed in the macroscopic world don't always extend reliably to the sub-microscopic domain. So the supervenience of the large on the small should not be beyond challenge - the fact that it has been very successful in some regimes does not necessarily entail that it is right in \emph{all} regimes.

A number of authors have argued that standard quantum mechanics offers some reason to be sceptical of reductionism in the ordinary sense. For example, \cite{Ismaelholism} argue that the phenomenon of non-separability pushes us towards a holist picture, and \cite{Koons2018-KOOHEI} argues that the infinite-dimensionality of thermal substances in quantum thermodynamics is evidence for a `hylomorphic' interpretation of quantum mechanics in which the physics at the large distances relevant to thermodynamics and  the physics at the much smaller distances relevant to particle physics stand in reciprocal relations of co-determination. Indeed, \cite{Miller2023-MILCRA-25} argues that even in the context of chemistry, quantum considerations seem to undermine petty reductionism. However, these discussions have largely focused on features of non-relativistic quantum mechanics; thus in this section,  I will consider what QFT has to tell us about the matter. I will argue that, over and above the points that have already been made in the literature about tensions between quantum mechanics and (petty) reductionism,  the structure of QFT in particular offers some additional reasons to be sceptical about petty reductionism in the regimes relevant to QFT. Note that in this section, I will be assuming that QFT as we currently understand it is at least \emph{compatible} with a non-reductionist picture; in the following section I will consider in more detail whether this compatibility assumption is justified.

\subsection{Infinities} 

As we saw in section \ref{intro}, quantum field theories are defined as a function of scale. And the versions of the theories defined at large distance scales (i.e. the scales at which we are currently able to probe them) are  well-defined, in the sense that they lead to finite predictions for scattering amplitudes and correlation coefficients - as indeed they must do if they are to be empirically adequate, since scattering amplitudes and correlation coefficients at large distance scales are empirically observable and naturally they are never observed to be infinite. But current evidence suggests that in the limit as we go towards very small distances, some of the quantum theories describing fields that exist in our actual world become ill-defined \citep{ButterfieldBouatta}: their coupling constants or masses become infinite, they no longer obey unitarity and they do not make finite predictions. And thus one may worry that at least the small-distance versions of these theories cannot be correctly describing reality, as many physicists and philosophers believe that there cannot be real infinities in nature. 

  The  usual realist response here is to assume that there exists  a physical cutoff below which QFT ceases to apply and instead some other physics takes its place, so quantum fields and their dynamics can be understood as playing a representative role only down to the cutoff \citep{wallace2001defence,cao2019conceptual}. That is to say, the usual response involves retaining reductionism whilst postulating that the very smallest things   on which everything else supervenes are not  quantum fields described by a small-distance Lagrangian - instead happenings at the smallest distance scales involve some new kind of physics which lies outside the framework of QFT.

However, there may be an alternative. Rather than replacing the small-distance QFT with something else, we could simply adopt a less literal attitude towards it - i.e. we could interpret the theory in such a way that  the Lagrangian parameters at small distances do not play a representative role, and thus the theory does not actually posit any physically real infinities. Now, in order to clearly distinguish this approach from operationalism, we would have to find a way of specifying a concrete ontology which allows us to divide the mathematical theory up into parts which are to be understood as playing a representative role and parts which are not so understood, ensuring that the `bare theory' falls into the latter category but it is not the case that \emph{everything} that is not observable falls into this category. And one obvious way to achieve this would be to adopt an ontology which postulates \emph{nothing} physically real at very small distance scales. Evidently a view of this kind would involve  a kind of non-reductionism: there would be physically real stuff  at larger distance scales but nothing physically real at the smallest distance scales, and therefore of course the physically real stuff at larger distances would not supervene on happenings at the smallest distances, since nothing at all would be happening at the smallest distances. Thus in principle a non-reductionist approach could help resolve the problem of infinities, by simply releasing the realist from the obligation to take these infinities literally.

\subsection{Naturalness} 

In the last fifty years a principle called `Naturalness' has become an important criterion guiding high energy physics research \citep{ROSALER2019118,pittphilsci15347}. Many different formulations of the Naturalness principle exists, but here I will follow Wallace in defining Naturalness as the requirement that the fundamental constants of nature should be selected from a probability distribution which is Natural, in the sense that it can be specified by some `\emph{not-ridiculously complicated function, relative to the uniform distribution}.' \citep{pittphilsci15757} 

\subsubsection{Is Naturalness Important? \label{why}}

Many physicists believe that some kind of Naturalness is a property we should expect or at least aspire to in our physical theories, though others   have opined that it is no more than an aesthetic preference which should not be taken seriously as a guide to fundamental physics. Hossenfelder, for example, points out that we have no real understanding of where the values of the physical constants come from and thus no reason to expect those values to reflect any particular probability distribution, so failures of Naturalness are not really a problem \citep{Hossenfelder_2019}.

However, Wallace argues that failures of `Naturalness' are in fact a problem - or at least, a problem for reductionists! Roughly speaking, the concern is that certain special choices of parameters can disrupt the usual relation between macroscopic dynamics and microscopic dynamics. That is, typically when we effect a reduction from macrodynamics to microdynamics, implicit in that reduction is the assumption that the initial conditions and fundamental constants are drawn from Natural probability distributions. This assumption is needed because different Natural distributions over the underlying parameters will lead to different values of the macroscopic parameters but the same qualitative form for the macroscopic dynamics, so `\emph{all of the information about the world's dynamics is encoded in the lowest-level dynamics, extractable from those dynamics through the assumption of Naturalness}'\citep{pittphilsci15757}, and this  makes formal reductions of the macrodynamics to the microdynamics comparatively straightforward to achieve. Whereas an un-Natural distribution over the initial condition and/or fundamental constants can potentially lead to a completely different qualitative form for the macroscopic dynamics, so with un-Natural distributions we can't derive the macrodynamics just from the microdynamics plus a generic assumption like Naturalness; the dynamics will be partly encoded in highly specific details of the initial state or values of parameters.   In particular, in the context of QFT, it is possible to choose un-Natural distributions with parameters that cancel out in a precise way such that a term which would otherwise have rapidly grown in size as we move to larger distance scales instead remains relatively small, so if we assume a Natural distribution we would expect that term to dominate the dynamics at larger distance scales, whereas with an un-Natural distribution it may not dominate at all, so we can get very different qualitative forms for the macroscopic dynamics even though the microscopic dynamics are the same in either case. 

Now, presumably in the actual world there is one actual set of physical constants, so while we may use probability distributions to characterise our knowledge of the constants or perhaps the process by which they are generated, when discussing the issue of ontological reduction we will want to refer to some specific set of values. So let us say that a particular choice of values for the constants is Natural provided that, in combination with the microdynamics and a Natural distribution over the initial conditions, it gives rise to the same qualitative form for the macroscopic dynamics as any Natural distribution\footnote{There do exist some theories, such as string theory and various multiverse proposals, which suggest that in fact there may be different values for the physical constants in different parts of the multiverse \citep{friedbook}. In this case, probability distributions over the constants may in fact  be understood to characterize an actual ensemble of universes with varying values for the constants. However, it is still the case in such theories that the behaviour we see in our own universe depends only on the actual values of the constants in our local region; so even in the multiverse scenario, it remains the case that the expected reduction of macroscopic dynamics to microscopic dynamics will work as expected only if the specific constants that hold for our own universe are `Natural.'}. Likewise, let us say that a specific initial condition is Natural provided that,  in combination with the microdynamics and a Natural distribution over the values of the fundamental constants, it gives rise to the same qualitative form for the macroscopic dynamics as any Natural distribution. That is to say, using a Natural distribution over parameters and initial conditions together with the correct microdynamics, as we usually do in attempting to understand the relation between microdynamics and macrodynamics, can be expected to lead to the correct macrodynamics if the actual values of the constants and initial state are Natural. 

But in fact, the values of some of the fundamental constants are known at least approximately, and according to current knowledge the values of the constants are \emph{not} Natural. There are at least two cases in which two of the constants appear to be very `fine-tuned' \citep{pittphilsci15757,pittphilsci15347} - which is to say, their values cancel out in a precise way such that a term which would normally have been expected to grow rapidly as the renormalization group flow moves towards large distances instead remain very small. In one case, the value of the cosmological constant at the cutoff scale ($\Lambda_0$) and another constant depending on vacuum fluctuations ($v$) appear to almost cancel out such that the cosmological constant at macroscopic scales ($\Lambda_M$) is much smaller than it would otherwise have been expected to be; in another case, the value of the bare Higgs mass ($m_H$) and quantum corrections coming from all the other standard model particles ($c$) appear to almost cancel out to result in the  observed value of the Higgs mass at the scales we are able to access, which again is much smaller than we would have expected based on calculations of the quantum corrections alone. So, insofar as reductionism depends on Naturalness, it appears that reductionism is failing at very small scales in the context of QFT. 

Now, clearly the issue Wallace raises is primarily a problem for \emph{methodological} reductionism rather than reductionism in any ontological sense. That is, even if the actual set of values for the parameters is  un-Natural, it could still be the case that everything going on at large distances supervenes on the goings-on at small distances; it would simply be   difficult for human scientists to \emph{write down} the form of the reduction from large distances to small distances, because the large distance dynamics couldn't be derived just from the small distance dynamics together with a generic assumption like Naturalness. Whereas the  success of ontological reductionism as a thesis about the makeup of our reality is, to some degree, independent of the facts about how easy it is for human scientists to actually arrive at formal reductions. However, as I have previously noted, much of the justification for believing in ontological reductionism in the first place comes from its spectacular \emph{methodological} success - so if it turns out that in the  regimes relevant to QFT methodological reductionism breaks down and we are no longer able to get good results by seeking to reduce the large to the small, that is surely at least some reason to think that we might be wrong about ontological reductionism in that regime as well. 

That said, one might worry that Wallace's account is to some degree exaggerating the threat to methodological reductionism. For after all, we \emph{have} arrived at these QFTs with apparently un-Natural choices of parameters, and we know how to derive high-level dynamics from them. So, whether or not the un-Natural parameter choices turn out to be right in the end, it is clearly not methodologically impossible for human scientists to arrive at  a reduction which involves un-Natural  choices of constants, even if this means we cannot  simply extract the macrodynamics from the microdynamics using a generic Naturalness assumption. Schematically it seems we have achieved this via a kind of bootstrapping: we started by assuming a Natural distribution, used this to establish microlaws, and then used those laws together with empirical tests to find out the actual values of the fundamental constants, leading to a more complex but still tractable reduction of macrodynamics to microdynamics. So it appears that the failure of Naturalness does not in and of itself block the application of methodological reductionism, if we allow more sophisticated methods such as the bootstrapping approach. 

But there is a more general methodological concern here. For empirical tests and bootstrapping can only take us so far: if the actual initial state and/or constants were too un-Natural this would undermine not just reductionism but the very practice of science. For example, if the initial state of the universe were not Natural, then there could be fine correlations built into it giving rise to a Boltzmann brain which happens to have all the memories that you now have, even though the events recorded in those memories never took place \citep{carroll2017boltzmann}. Evidently empirical tests will do little to help prove that this is not the case, for any empirical results we might obtain could equally well be fabricated. So in order to make sense of the practice of science, it seems we need to make some kind of Naturalness assumption in order to rule out maverick possibilities like Boltzmann brains. But then the observation that some of the fundamental constants appear to be  fine-tuned puts us  in the awkward position of having to insist as a kind of a priori methodological assumption that the initial condition and constants are almost entirely Natural, but there are just a few deviations from Naturalness which manifest entirely in effects that we are able to detect empirically but which never go so far as to  underminine the methods we typically use for empirical testing. Of course, it could be the case that the world is that way, but one might feel there is something epistemically irrational about \emph{believing} it to be that way: after all, from another point of view one might be inclined to react to the evidence of Naturalness violations by concluding that they suggest the existence of  much more widespread Naturalness violations which should lead us to conclude that our means of empirical enquiry may be unreliable. 

So although the existence of a few small failures of Naturalness do not necessarily  count as evidence against either methodological or ontological reductionism, they do appear to pose a more general problem, since they potentially undermine our understanding of the rationality of scientific enquiry.   Worries of this kind therefore give us clear motivations   to try to preserve Naturalness for \emph{epistemic} reasons.

\subsubsection{How can we preserve Naturalness?}

If it is accepted that preserving Naturalness is important, what are the options for doing so?  Other than holding on to hope that  the empirical evidence for the fine-tuned values will one day be overturned or significantly reinterpreted,  the most obvious way out would be to say that these problematic fine-tuned constants (e.g. $\Lambda_0$ and $v$ or $m_H$ and $c$) are not, in fact, fundamental - there is some new physics to be discovered which will explain how their values come to be so close together as a consequence of their jointly arising from some other more fundamental constant(s). This would amount to saying that the Naturalness assumption is right, but is simply being applied in the wrong place: we should expect the constants which are truly fundamental to be Natural, but we shouldn't in general expect non-fundamental constants to be Natural, because they will usually stand in various mathematical relations as a consequence of the ways in which they are related to the real fundamental constants. So the question about where exactly the Naturalness assumption ought to be applied provides a concrete example of a case in which questions about reductionism come to have real scientific significance: what we take to be fundamental has a real impact on the decisions we will make here about what counts as a real problem for physics and the direction in which future research should go. 

In the spirit of petty reductionism, the most obvious route here would be to postulate some new physics at \emph{smaller} scales - so for example we could postulate some new constant which is relevant at scales smaller than the scale at which $\Lambda_0$ and $c$ are defined, and suppose that both  $\Lambda_0$ and $c$  are directly related to this more fundamental constant, via something like a renormalization transformation or some other kind of coarse-graining, in a way which explains the closeness of their values. But   there would be something a little odd about this picture, for it seems to involve a `double convergence,' in which there is a single fundamental constant at an extremely small scale, which through some mechanism (perhaps a version of coarse-graining, or a renormalization transformation) is transformed into two apparently distinct quantities - $\Lambda_0$ and $c$ -  at an intermediate scale, which in turn are transformed via the renormalization transformation into a single quantity $\Lambda_M$ at larger scales. Moreover the fact that the intermediate quantities are related in this way to the fundamental constant would explain why they are close in value, but wouldn't necessarily explain why they are close in value in just the right way to cancel out to produce that low value for $\Lambda_M$ at larger scales. This double convergence isn't impossible, of course, but one might worry that there is still something a little conspiratorial about it, or even that it is still fine-tuned and hence doesn't avoid the original problem. And moreover,  similar concerns would surely apply to any other kind of reductionist mechanism that we might posit to explain the closeness of the intermediate values - it's unclear that we can reasonably expect to get this specific value for $\Lambda_M$ without fine-tuning whatever is going on at scales smaller than $\Lambda_0$ and $c$.

And indeed, is this not unnecessary? We already have a quantity which is directly related to both $\Lambda_0$ and $c$ - $\Lambda_M$! The renormalization transformation is reversible, after all, or at least some versions of it are: and thus, just as we can derive from the fact that $\Lambda_0$ and $c$ are very close in value  that $\Lambda_M$ must  be small, so too can we derive from the fact that $\Lambda_M$ is small that $\Lambda_0$ and $c$ must be very close in value. That is, if we start  from a fixed, very small value for the cosmological constant at macroscopic scales, $\Lambda_M$. then we can immediately derive that $\Lambda_0$ and $c$  must be `fine-tuned' in exactly the way that they are, since they must cancel out to give the predefined small value for the cosmological constant. This leads naturally to the conjecture that perhaps in this scenario we should consider $\Lambda_M$ to be `fundamental,' i.e. ontologically prior to $\Lambda_0$ and $c$. Then the appropriate way of applying the Naturalness assumption here would be to apply it to $\Lambda_M$ together with various other fundamental constants, excluding $\Lambda_0$ and $c$ - so with $\Lambda_0$ and $c$  not  in the domain of the Naturalness assumption at all, we might be able to maintain Naturalness as applied to fundamental constants. Obviously, much here would depend on what else was included in the set of fundamental constants - nothing would be gained by this strategy if we found we could only make it work by including some other quantity which looks fine-tuned relative to the value of $\Lambda_M$ - but nonetheless, this example at least suggests that we don't necessarily need to solve Naturalness problems by appeal to brand new physics. Reinterpreting the physics we already have in a less reductionist way could potentially also solve the problem, at a lower ontological cost. 

To summarize,  one natural way to avoid the epistemic problems discussed in section \ref{why}  would be to  apply the Naturalness assumption not to physics defined at the smallest scales but to physics defined at some larger scale. Thus, insofar as it makes sense to expect the Naturalness assumption to apply at the most `fundamental' level, this line of reasoning gives us reason to consider moving away from ontological reductionism. It is in this sense that failures of Naturalness give us reason to take seriously a non-reductionist view of QFT - not because failures of Naturalness are direct evidence for failures of ontological reductionism, but because failures of Naturalness lead to a severe epistemic problem which can potentially be defused by adopting a non-reductionist view.

 \subsection{Finite supervenience base \label{finite}}

 A QFT is referred to as `non-renormalizable' if the bare Lagrangian at the smallest possible scales, as defined in the continuum approach, requires an infinite number of terms \citep{ButterfieldBouatta}. Non-renormalizable QFTs would seem to lead to difficulties for reductionists, or at least for  reductionists of the \emph{foundationalist} persuasion,  as it seems natural for those who believe that there is some fundamental level of reality on which everything else in the physical world supervenes to also believe that this supervenience base should be \emph{finite}, i.e. specifiable in a closed form. Perhaps there may be no closed-form description of the higher-level facts supervening on the basis, but one would naturally hope for a finitely specifiable base level, or else it's unclear what is really gained by going to this supposedly most fundamental level.

 Now, as a matter of fact most of the QFTs describing fields in our actual world (e.g. quantum electrodynamics, quantum flavour dynamics, and quantum chromodynamics) do in fact appear to be renormalizable \citep{ButterfieldBouatta, berestetskii2012quantum,Fritzsch1977,Greiner2007}. However, this does not mean that reductionism is safe. For \cite{ButterfieldBouatta} note that there is a straightforward explanation for the renormalizability of these theories: `\emph{for any such theory—
“with whatever high-energy behaviour, e.g. non-renormalizable terms, you like”—the
non-renormalizable terms dwindle into insignificance as energies become lower and
length-scales larger}' because `\emph{the physical coupling constant
for non-renormalizable terms shrinks.}' In this context, we say that `irrelevant' terms in the Lagrangian  have coupling coefficients which are large at small distances but go to zero in the limit as the distance becomes large, so they will not appear in the large-distance theory but they have to be added in as we go to small distances; whereas  `relevant' terms in the Lagrangian have  coupling coefficients which are large at large distances but go to zero in the limit as the distance becomes small, so they will not appear in the \emph{small}-distance theory but they have to be added in as we go to large distances \citep{pittphilsci21666}. So the comments of \cite{ButterfieldBouatta} amount to the observation that a theory which appears  renormalizable based on the relevant terms present in its large-distance formulation can still in principle have an infinite number of irrelevant terms appearing at small distances, and thus our current evidence does not rule out the possibility that QFTs describing our actual world may fail to be finitely specifiable at the smallest distances.

Moreover, it can be shown that as a consequence of structural features of a QFT, there can only be a small number of physically possible relevant terms in such a theory, whereas there will be an infinite number of physically possible irrelevant terms \citep{peskin,skinlec}. So the worry that in realistic QFTs the small-distance theory may fail to be finitely specifiable is not just a wild speculation - the mathematical structure of the theory gives us reason to take this seriously. And thus the state of our current understanding makes QFT look quite unique as compared to prior theories. Throughout the history of science we have usually found that  that our theoretical descriptions become in some sense simpler as we move to smaller distances - this is the case in paradigmatic examples of theoretical reduction, such as the reduction of a large number of different chemical elements to  arrangements of protons, neutrons and electrons, and indeed the general pattern of increasing simplicity as we go to smaller distances is surely one of the main reasons why reductionism as a methodology is so appealing and so successful. So even though the current evidence does not prove conclusively that there is no finite supervenience basis at very small scales, nonetheless even the fact that there appear to be  reasons in QFT to take this possibility seriously is novel, and thus we should   consider what would follow if this possibility in fact holds. 

Now, the standard reductionist response to this problem is to suggest that there exists some physical cutoff below which the effective field theory framework no longer applies, so the apparent problem just comes from extrapolating the theory beyond its domain of validity. However, an alternative way of addressing the problem might involve the proposal that we have the  direction of supervenience  the wrong way round. For although there is a question mark over whether the small-distance Lagrangians are finitely-specifiable,  both our existing empirical evidence and our understanding of the theoretical structure confirms that the number of relevant terms in realistic Lagrangians is small \citep{peskin,skinlec}, so we can reliably expect  that large-distance Lagrangians will have a finite number of terms in some appropriate limit. At the scales we ourselves  probe the irrelevant terms presumably do not have coefficients of exactly zero - they simply have coefficients which are close enough to zero that these terms can be neglected -  but in the infinite limit these coefficients will go to zero and thus in that limit it seems plausible to say that the theory  is  finitely specifiable. Of course, the infinite limit is usually regarded as an unrealistic idealization; but if the only scale at which the theory can be finitely specifiable is in the infinite limit, I think there is at least a prima facie case for regarding this limit as the  most fundamental statement of the theory.  Then  to find the theory at any smaller distance we would in principle just iterate the renormalization transformation, adding new terms as needed as we reach smaller scales. Since the limit is a fixed point,  the initial transformation away from the fixed point would not be one-to-one, so we would potentially have to specify some means of breaking the symmetry at the fixed point, but this seems possible at least in theory. Since there is no particular reason to think we must be able to specify  everything that supervenes on our supervenience basis in a closed form, it would then no longer matter if in fact the small-distance theory is not finitely specifiable.  

To be clear, my intention here is not to suggest   that because we are currently unable to determine whether or not the small-distance theory is finitely-specifiable, we should conclude that the large-distance theory must be fundamental: that would be to mistake a limitation of our present epistemic access for a fact about ontology. I simply mean to suggest that  as long as there are reasons to take seriously the possibility that the small-distance theory is not finitely specifiable, this amounts to a further indication that   non-reductionism \emph{may} have some part to play in making conceptual sense of QFT (although it is only one alternative amongst various possible solutions to the problem, and the cutoff approach also remains a reasonable option).  These considerations, in combination with some of the other arguments of this section, contribute to the overall picture suggesting that petty reductionism should not be taken for granted in the context of QFT.

\subsection{Divisibility} 

The traditional reductionist picture is based on the idea that we can break nature down into its `smallest pieces,' each having a characteristic set of intrinsic properties and behaving autonomously except during interactions, and thus we can infer what higher-level behaviour will emerge from the complex interactions of all of these distinct, autonomous pieces - just as  we can understand the functioning of a complex machine once we know the properties of each of its individual parts and the way in which they were put together. So for example, we can understand the vast array of chemical compounds in the world by understanding how they are all composed from different combinations of atoms, each atom having fixed intrinsic properties such as charge and mass. 

Because the traditional reductionist picture leans heavily on the notion of distinct autonomous pieces, the phenomenon of nonseparability in quantum mechanics seems to threaten traditional reductionism, insofar as it gives us reason to doubt that the universe can really be divided into such autonomous pieces. \cite{Ismaelholism} have made this point in the context of standard quantum mechanics - they argue that  `\emph{(quantum mechanics) allows spatiotemporally separated entities to have states that cannot be fully specified without reference to each other}' and thus we should be `\emph{viewing nonseparable entities (such as Alice and Bob) in a holistic light, as scattered reflections of a more unified underlying reality.'} 

And in fact, an even more compelling version of this argument can be made in the context of QFT. For in non-relativistic QM it typically seems possible at least conceptually to think of the parts of a composite system as distinct from one another, even if nonseparability prevents us from defining their states or dynamics  autonomously. But in the context of QFT it seems difficult to even conceptualize what it would mean to break composite  systems into distinct pieces, because this would involve `\emph{trying to characterize one aspect of an interacting many-field system in a way that is comparatively simple and independent of the rest of the system}' \citep{ButterfieldBouatta} and we can't  meaningfully take the fields apart in this way. For example, \cite{ButterfieldBouatta} explain the mass dependence of particles in QFT by analogy with the change in apparent mass of a ball when it is placed in water versus air, but then note that QFT is in fact significantly different from this kind of case, because we can take the ball out of the water, but in QFT `\emph{we cannot take the system out of the field of force, which is all-pervasive, though varying in strength from place to place.}'

Now one might think that this is just a practical, experimental barrier and that nonetheless each system in QFT still has well-defined intrinsic properties, such as charge and mass, which determine its behaviour in any given field. This would entail that even if we can't actually take the systems out of the field, nonetheless we should be able to formulate counterfactuals about what we would observe if we could take such systems out of the field, since that leaves us with the intrinsic properties.   But in fact, the barrier here is not just a practical one, because a system in QFT is nothing more than an excitation in a field, and interactions are represented by other fields, which means that even in theory the idea of taking the system out of the field isn't really coherent: `\emph{the physical system is an interacting many-field system, so that it makes little sense to conceptually detach one of the fields from the others}' \citep{ButterfieldBouatta}. Thus in a sense what QFT reveals is that as we go to smaller and smaller scales, the world becomes more and more non-separable, and thus more and more resistant to being divided up into autonomous pieces, so that  the interpretation in terms of some kind of holism becomes ever more compelling. In particular, in QFT it would seem that there is no   reason to expect the individual systems, or constituents of quantum fields, to have  well-defined intrinsic properties which determine what would occur if we could take them out of the field of force - and thus it is perhaps not so surprising that we get nonsensical answers like `infinity' when we try to calculate intrinsic properties like mass and coupling constants for quantum fields.

Of course, one might accept this argument and yet contend that non-separability and holism are not necessarily incompatible with ontological reductionism - one might imagine that the reductionist can still make sense of the notion that large-distance effects  supervene on small-distance phenomena even if those small-distance phenomena can't be understood in terms of individual autonomous parts in the traditional way. But nonetheless, moving towards some kind of holism is already a significant departure from the intuitively compelling picture of a world in which all large-distance objects and structures are simply composed out of distinct autonomous parts, so it certainly isn't good news for ontological reductionism that as we go towards the level of reality where the ontological reductionist wishes to site their fundamental supervenience basis, it appears to get more and more difficult to separate nature into autonomous parts. So insofar as the structure of QFT can be interpreted as pointing towards a kind of holism, this also provides additional reason   to take seriously the possibility that ontological reductionism may fail in regimes relevant to QFT. 
 
\section{Is Non-Reductionism Compatible with the Structure of QFT?}

The issues  raised in the previous section are suggestive, but of course none of them  conclusively proves the failure of reductionism in QFT, since there are possible options other than non-reductionism  to address theses issues. In particular, many physicists consider that all of these problems can all be addressed by employing the effective approach with a physical cutoff, in which case many of the problems described in the previous section can potentially be thought of not as representing actual pathologies in the theory but as arising simply from the fact that we are moving outside of the domain of validity of the effective theory. For example,   we don't have to worry if infinities appear at scales smaller than the cutoff since those infinities will not be real; we can  hope that new physics below the cutoff (or perhaps at a higher scale) will provide a mechanism to explain why Naturalness fails for the parameter values currently known; we can hope that ultimately the physics below the cutoff will be finitely specifiable and simpler than the large-distance QFTs; and maybe we can even hope that whatever lies below the cutoff will be compatible with being broken down into separate autonomous pieces, as in the traditional reductionist picture. 
 
Now, it is not my intention to argue against the cutoff approach in this paper. There are indeed good reasons to believe that quantum field theories should stop working below some scale - for example, some approaches to quantum gravity suggest that spacetime is likely to be fundamentally discrete, which would explain why there must be a cutoff \citep{cc}.   However, I think it is  nonetheless worthwhile to explore whether we could find some partial hints to the solutions to the problems described above  within the existing theory, rather than deferring these problems to some as yet undiscovered physics. And I take it that the arguments of the previous section offer some reason to think that  adopting a form of non-reductionism may potentially offer an alternative solution to these problems which is independent of future developments in physics.  Thus if the non-reductionist view can be shown to be a viable alternative, it should be taken into account as we consider our options for responding to these problems. 

Thus in this section, I will consider in more detail whether or not  non-reductionism in the ontological sense is in fact a viable alternative. That is, given what we know about the mathematical structure of QFT, is a non-reductionist interpretation of it technically possible? In particular, if the actual scale transformations instantiated by physics in our world  were many-to-one towards larger distances, then ontological non-reductionism would appear to be defeated on purely technical grounds: for such a mapping would entail that the large-distance properties cannot differ without differences in the small-distance properties, whereas the small-distance properties  can indeed differ without differences in the large-distance properties, and therefore we would be presumably be obliged to say that the large-distance properties  supervene on the small-distance properties and not vice versa. 

However,  we  saw in section \ref{intro} that there are good reasons to think that outside of fixed points the actual scale transformations instantiated by physics in our world may be   one-to-one. And thus the relation between the large-distance physics and the small-distance physics encoded in the renormalization group flow  may actually be fairly symmetric: the large-distance physics determines the small-distance physics and vice versa. Of course, this is only true if we are working with the \emph{complete} large-distance physics - as noted in section \ref{intro}, if the large-distance Lagrangian is missing some terms or does not include fields which become important at small-distances, or if we start from an approximate large-distance Lagrangian with some coupling coefficients set to zero, then applying the renormalization transformations to the large-distance Lagrangian will not produce the right small-distance physics. But from the point of view of ontological reductionism what matters is not whether the transformation can be practically effected by human scientists but whether the mathematical structure of the theory admits a one-to-one relationship at least in principle, and it does appear that in the ideal case where we start from a correct and complete large-distance Lagrangian, then the relationship may indeed by one-to-one. Thus based on current understanding of the mathematical structure of the theory it does not appear that we are forced to say the large-distance physics supervenes on the small-distance physics and not vice versa; in some sense it looks as if the small and large distance physics appear to have similarly good claims to fundamentality based purely on the structure of the physics.

I will now examine four further approaches the reductionist might take to argue that the mathematical structure of QFT supports ontological reductionism, arguing that none them is fully convincing. Note that the goal of this section is not offer more arguments in favour of the non-reductionist approach, but simply to show that such a picture is not \emph{ruled out} by the theoretical structure of QFT. This is important since the considerations of the previous section cannot be interpreted as evidence in favour of a non-reductionist picture if it turns out that the structure of the theory in fact compels us to adopt a reductionist picture, so the task of this section is simply to show that in fact the structure does not so compel us.

\subsection{Additional Terms}

One possible argument for ontological reductionism in QFT could involve contending that although the scale transformations obtained by renormalization group methods may appear to be one-to-one, in fact this fails to be the case once we take into account the possibility that we may have to add additional terms into the Lagrangian as we go to smaller distances, since terms which had coefficients of zero in our original Lagrangian may come to have non-zero coefficients under the renormalization group flow. The appearance of new terms may appear to show  that the small-distance physics is not fully determined by the large-distance physics after all, which would undermine the notion that there is a symmetric relation between physics at large and small distances. 

However, this argument does not succeed, since the same thing can occur as we move from small distances to larger distances. As discussed in section \ref{finite}, typical QFTs will include both relevant terms which disappear at large distances, and irrelevant terms which disappear at small distances (as well as marginal terms which belong to neither category).  Thus regardless of whether we are going from small distances to large distances or vice versa, it is possible for new terms to appear in the Lagrangian. 

It is true that the mathematical structure of a QFT is such that there are always many more irrelevant terms than relevant terms \citep{peskin,skinlec}, meaning that the problem of new terms appearing is in some sense worse  as we go from large to small distances rather than vice versa. And certainly the large number of irrelevant terms would present complications in actually writing down a transformation from large distances to small distances, so from the point of view of practical applications it might be easier for human scientists to derive large-distance physics from small-distance physics rather than vice versa. However, the number of such terms does not affect the argument in principle: if it is the case that the appearance of new terms as we move from one scale to another means that the original scale cannot be fundamental, then the fact that new terms can appear regardless of whether we are scaling up or down in a QFT would entail that in such theories small nor the large distance theories can be regarded as fundamental. Thus  even when we take the appearance of new terms  into account, the mathematical structure of the renormalization group doesn't compel us to say that  large-scale effects must supervene on the small-distance happenings rather than vice versa.

\subsection{Fixed Points}

Another possible argument  for ontological reductionism in QFT might be based on the existence of fixed points. For the definition of fundamentality I have been using in this article means that physics at a given scale $S$ can be fundamental only if it is possible for all physics at  other scales to supervene on the physics at scale $S$. The reductionist might therefore suggest that the physics at a fixed point of the renormalization group flow cannot be `fundamental,' since the map going to a fixed point is many-to-one and thus a number of different Lagrangians at $L$ will be compatible with the same Lagrangian at the fixed point, meaning that  the physics at some other scales $L$ cannot supervene on the physics at a fixed point. Thus the reductionist might argue that since the renormalization group flow can have fixed points at large distances, it's impossible that the physics are large distances is fundamental. 

 As a first response, note that fixed points arise when we take the infinite limit which fully removes the cutoff \citep{ButterfieldBouatta}. If one regards this limit as unrealisable or unphysical,  fixed points are also unphysical and thus we can be assured that the relations between physics at different scales will always be one-to-one. In this case, the existence of fixed points would not imply anything about the fundamentality of physics at different scales. 
 
 But even if one does not accept that fixed points are unphysical, the existence of fixed points would not clearly favour either the reductionist or the non-reductionist picture. This is because  fixed points can occur either at large distances (IR fixed points) or small distances (UV fixed points), and indeed there is  evidence that both kinds of fixed points occur in real field theories in our actual universe: it is known that QED has an IR (quasi-)fixed point \citep{PENDLETON1981291}, whereas QCD is currently thought to have a UV fixed point \citep{Bethke_2007}. So regardless of what one thinks of the idea  that fixed points cannot be fundamental, there does not appear to be any general argument from the existence of fixed points to the claim that the small-distance physics must be fundamental. Perhaps we might be tempted to say that in theories with IR fixed points, the small-distance physics is fundamental, while in theories with UV fixed points, the large-distance physics is fundamental - but since both types of theories are realised in our actual world, and the entities that they posit interact with one another, this would seem to destroy any hope of positing a single, fundamental level of reality from which all else emerges - at small scales we will have some `fundamental' entities interacting with some `non-fundamental' entities, and likewise at large scales, so there can be no unique `fundamental' description.

\subsection{Dynamics Versus States \label{states}} 

Another possible argument for ontological reductionism in QFT  might be based on scale transformations for \emph{states}. We have seen that the dynamics of quantum field theories and the scale transformations between dynamics at different scales does not  compel us to accept that large-distance physics must supervene on small-distance physics. However, one might think that we can still arrive at this conclusion by focusing on the states. After all, at least in the effective field theory approach to renormalization, the mathematical definition of the renormalization transformation assumes that there is a many-to-one relation between the specific small-distance state and the larger distance, more coarse-grained state  - the whole point of the transformation is that we are integrating out high energy degrees of freedom, thus losing some information about the small-distance state. That is, states at larger distances supervene on the smaller-scale states, but not vice versa, and so it appears as if the most fundamental description of the state of a system must be given at the smallest possible scales. This  conforms with our usual intuitions about the relations between states at different scales, as instantiated for example in the relation between thermodynamics and statistical mechanics, where we usually take it that a single thermodynamic state may be compatible with  a large number of possible underlying microstates.

However, care must be taken with this line of reasoning. First, remember that the definition of the renormalization group transformation which involves `integrating out degrees of freedom' is not the only possible way to arrive at that transformation: as we have seen, the continuum approach uses an alternative method which does not involve integrating out degrees of freedom, and still arrives at what appears to be a correct result. So it is not clear that we have to take the picture involving `integrating out degrees of freedom' literally. Furthermore, the very fact that we integrate out degrees of freedom to arrive at our predictions for the results of the measurements we can perform entails that most specific details of those degrees of freedom are not relevant to our predictions: nearly everything that we predict depends not on the specific states of the underlying degrees of freedom but on the underlying \emph{dynamics} and their relation with larger-distance dynamics. And insofar as states are relevant, what matters is not the specific short-distance state, but very general features of that state. In particular, the EFT methodology relies on a perturbative expansion \citep{peskin}, so the short-distance state needs to be well-behaved enough for the expansion to be valid: and of course this assumption rules out some class of possible short-distance states, but certainly it does not commit us to any particular short-distance state, and nor does it appear to depend on the claim that there \emph{exists} some particular short-distance state. Thus it isn't at all clear that the  methodology of quantum field theory  commits us to taking all small-distance degrees of freedom literally. 
 
Additionally, it is important to remember that the small-distance state is not a microstate in the classical sense - it is not a classical configuration of particles or of fields. Rather it is a   \emph{quantum state} of a field, and the ontological status of the quantum state remains an open question even at the level of  non-relativistic quantum mechanics. Thus, given the big question mark over the right way to think about the quantum state, perhaps it should not be taken for granted that there is really a many-to-one map from low distance states of a quantum field to larger distance  states of a quantum field. 

To underline this point, first recall that in standard  non-relativistic quantum mechanics, when we know that a system has been prepared in a pure state from the set $\{ \psi_x\}$ with probability $p(i)$ for state $\psi_i$, we may represent the   state of the system as a mixed state: $\rho = \sum_i p(i) \psi_i dx$ where $p(i)$ is the probability we ascribe to the system being in the pure state $\psi_i$.  And standard quantum mechanics assures us that if this probabilistic preparation is performed a large number of times, the relative frequencies arising from experiments performed on these systems will be exactly the same as the relative frequencies that would result from a set of systems all prepared identically in a single   mixed state  $\rho$, rather than in a probabilistically selected pure state. That is to say, there is no measurement which will distinguish between the proper mixed state arising from the probabilistic preparation and corresponding scenario where there is just one preparation producing a single mixed state. 

Now, the coarse-graining transformation associated with the renormalization group flow in QFT is typically designed in such a way that a pure state at distance scale $L'$ on a Hilbert space $H'$ will be taken to another pure state at a larger distance scale $L$, on some other Hilbert space $H$. But now suppose that you are observing a system at the distance scale $L$ and wondering about its state at the distance scale $L'$. You will not be able to infer the exact  state of the system at the scale $L'$ from your knowledge of the state $\psi$ at scale $L$, since the coarse-graining transformation acting on states is many-to-one. You will only be able to infer that the state at scale $L'$ belongs to some region $S$ of the Hilbert space $H'$, corresponding to all of the states on $H'$ which would be taken to $\psi$ under the appropriate coarse-graining transformation. So you might perhaps encode your knowledge of the state at scale $L'$ as a generalized verison of a mixed state $\rho' = \int_S \rho dx$, where we integrate over all states $\rho$ belonging to the region $S$ using some measure $dx$. And  the mathematical equivalence between proper and improper mixed states suggests that, as long as you observe only at the distance scale $L$ and refrain from probing the distance scale $L'$, the behaviour you observe will be compatible with the system \emph{really} being in the mixed state $\rho$, rather than in some unknown state in the region $S$. Indeed,  $\rho$ itself will presumably be in the region $S$, since if a set of states are all taken to a single coarse-grained state under a coarse-graining transformation, then for most natural coarse-graining transformations we would naturally expect that a convex mixture of them would also be taken to that state. Certainly, we know already from the fact that the renormalization group transformation does not depend on the specific state  (provided that the high-energy degrees of freedom are sufficiently well-behaved) that the \emph{dynamics} you observe at scale $L$ will be relatively insensitive to the specific state at a given scale. So in general the observed dynamics won't be in contradiction with the hypothesis that the state at scale $L'$ is the unique state $\rho$, rather than some unknown state in the region $S$.  

So under these circumstances, are we still justified in insisting that all along a quantum field is \emph{really} in some specific unknown state in the region $S$, rather than in the mixed state $\rho$ or some appropriate generalization of the mixed state concept? At the very least, this assertion seems highly interpretation-dependent: proponents of the Everett interpretation would likely espouse it, whereas proponents of views which take the quantum state less literally, such as QBism \citep{QBismintro} or the Bell flash ontology \citep{Gisin2013,Tumulka_2006}, might be less inclined to espouse it.  Of course, it may well be that every time we actually probe the system at the distance scale $L'$ and perform tomographic measurements on it we do find it in some specific state, but given the ontological murkiness of the quantum state, it's unclear that such results can be interpreted as telling us that it really was in that state all along,  for after all one of the key lessons of quantum theory is that measurements are not just passive observations; they can sometimes effect substantial changes upon the systems being measured. It is true that in non-relativistic quantum mechanics  there exist certain sorts of measurements, including weak measurements \citep{Kastner_2017,PhysRevLett.60.1351} and non-selective measurements \citep{PhysRevLett.110.260502}, which arguably can be interpreted as giving us information about the state without  changing it in a significant way; however, given that measurement theory in QFT remains a difficult topic \citep{Papageorgiou2024-PAPETI}, it isn't clear that such things generalize to QFT, or that they can be interpreted in the same way in QFT. 

Therefore at least at present it should not be taken for granted that the supposed `coarse-graining' transformation from small distances to larger ones in QFT is literally a many-to-one map on states. Certainly the way in which it is formally implemented \emph{looks} as if we are integrating out degrees of freedom, but the formal implementation in and of itself does not necessarily commit us to the view that what actually exists at small scales is some specific unknown state, as opposed to some kind of ontological mixture of all of the small distance states compatible with the large-distance state. And thus it does not appear that we are forced to  think of the renormalization transformation as  losing information about states, in which case we are also not forced to accept that large-distance descriptions must supervene on small-distance descriptions in QFT.

 .

\subsection{Condensed Matter}

Another possible argument for ontological reductionism in QFT might be based on the well-publicized analogy between  renormalization in QFT and renormalization in condensed matter \citep{kadanoff2010theories,RevModPhys.55.583,KADANOFF201322}. After all, in  the condensed matter case, we have a clear ontology in mind - a lattice with an individual quantum system at each lattice site - and at least in a \emph{classical} condensed matter system, we know that many different configurations of the individual lattice systems can give rise to the same macrostate of the entire condensed matter system, so the map from small-distance states to large-distance states is clearly many-to-one,  and thus there is surely little doubt that the macrostate supervenes on the small-distance configuration. And since  renormalization in QFT has a number of formal similarities to renormalization in condensed matter systems \citep{pittphilsci14591}, it seems natural to assume that the reasons for its success are the same in both cases, and thus to imagine that in QFT it must still be the case that the larger-distance behaviour supervenes on the smaller-distance behaviour.

However, some caution is necessary here. In the case of  condensed matter systems (or at least, \emph{classical} condensed matter systems) we have very strong empirical and theoretical reasons to believe that for any given coarse-grained state of the system, all of the various fine-grained states which are compatible with that coarse-grained state do actually occur in real systems. For certain kinds of condensed matter systems we can use microscopes to actually observe fine-grained states, and there are various kinds of experimental observations we can make to  learn something about which one of the various microstates compatible with the macrostate is the actual microstate of the system. So it wouldn't be plausible in this context to suppose that each coarse-grained state really corresponds to a single `mixed' fine-grained state. But in the context of QFT both the experimental and the theoretical situation is different: experimentally we have no ability to probe the individual states at low distances, and theoretically we don't have a well-defined description of what could be going on at the smallest relevant scales, so it's certainly not clear that we are still obliged to take the set of underlying microstates so literally.

Furthermore,  although there are formal, mathematical similarities between renormalization in condensed matter systems and QFT, there are disanalogies as well. In particular,  \cite{pittphilsci14591}  gives a detailed account of the mapping between the renormalization transformation in condensed-matter and the analogous transformation $T_1$ as used in continuum QFT (as defined in section \ref{renormalization}). Although the mathematics of the condensed matter transformation can be used to derive the QFT transformation, Fraser notes a number of key disanalogies; and most importantly for our purposes, she emphasizes differences in the modal status of the different transformations. Specifically, the condensed matter transformation relates various possible states that the same physical system could have at different times; the flow it describes can therefore be understood as a process of evolution in time (e.g. evolution towards a critical point) and can be empirically observed. Whereas we saw in section \ref{intro}  that the QFT transformation $T_1$ relates hypothetical descriptions of the system with different possible cutoff lengths, so the flow it describes is not a real physical process and can't be empirically observed. Thus   the formal similarity between these transformations does not mean   that the two processes or the systems in which we apply them are \emph{physically} similar in all possible ways - the transformation $T_1$ is not a process that can literally occur, it is just a convenient formal manipulation, so in and of itself it can't tell us much about the physical nature of the QFT systems in which it is performed. 

On this basis, one might be more inclined to make a comparison between the renormalization transformation in condensed matter and the QFT transformation $T_2$, which \emph{does} refer to something we can empirically observe - the running of the coupling constants with scale has indeed been empirically tested. Although the analogy with condensed matter was originally used in the construction of the transformation $T_1$, one might argue that the  success of this analogy is really a consequence of the fact that the condensed matter transformation is quite similar  to transformation $T_2$, which we know is closely related to $T_1$. Indeed, if one interprets  the effective approach by postulating that there exists a real physical cutoff at some small distance scale, the analogy looks even closer, because condensed matter systems really do have a physical cutoff at small distances (the continuum description of the condensed matter system ceases to apply once we have zoomed in so far that the scale is smaller than the lattice spacing) and  it is sometimes suggested that this fact provides some evidence in favour of the cutoff interpretation of effective QFT.

However, even in this case there are important disanalogies. For the  condensed matter transformation describes a set of states which can be instantiated at different times,  so we can observe one and the same system moving between these states, whereas the QFT transformation seems to describe a single system at a single time at different scales. Yet  we can only observe a system at one scale at any one time, so we can't really observe one and the same QFT system at different points along the renormalization flow: rather we observe an ensemble of such systems and probe them all at different scales. So the interpretation of the QFT transformation necessarily involves more steps then the interpretation of the condensed matter transformation, since the QFT case requires us to invoke a whole ensemble whereas the condensed matter transformation can be thought of as describing a succession of states experienced by a single system. 

Moreover, the interpretative issues around `states' that we described for the QFT case in section \label{states} don't come up in the same way in condensed matter. For the condensed matter system undoubtedly has a sequence of distinct time-indexed states related by the renormalization group flow,  whereas it's not  clear that we can say  the QFT system has a sequence of \emph{distinct} states related by the renormalization group flow - those states can only be indexed by scale, rather than time, and one might think that in some sense the  states at smaller distances compose the states at larger distances, so their distinctness is in question. Moreover, in practice the relevant `scale' is usually set by an observer who probes the system at some scale, so indexing states by scale can't be done in an interpretation-neutral way - it requires us to make some assumptions about observers and the act of measurement,  none of which are needed if the states can be indexed by times instead.

This distinction is also relevant to the idea that the existence of a real  cutoff in condensed matter provides some evidence for the cutoff interpretation of effective QFTs. For in the condensed matter case, the cutoff is in units of \emph{distance} but the renormalization group flow can be thought of as occurring along the \emph{temporal} axis, with the correlation length getting larger and thus the effect of the cutoff becoming less important as we approach the critical point. Whereas in the case of effective QFT with a physical cutoff, the cutoff is in units of distance and the renormalization group flow also occurs along the axis of distance, so again the effect of the cutoff becomes less important as we go to larger scales. While it is true that there seems to be some common basis to the two effects, with cutoffs becoming less important as we move to larger distances, nonetheless the fact that in the QFT case the putative cutoff is on the same axis as the trajectory of the flow whereas in the condensed matter case the putative cutoff can be thought of as being on a completely different axis should at least raise questions about whether the putative QFT cutoff can really be understood as playing the same role as the condensed matter cutoff. At the very least, this distinction indicates that the putative cutoff in QFT clearly can't play \emph{exactly} the same role as the cutoff in condensed matter, and this  gives some grounds to think that the formal similarity between the procedures may have its origins in some other kind of physical similarity, rather than the existence of a literal cutoff in both cases. So although the analogy to condensed matter has been very fruitful and will no doubt continue to yield insights, it shouldn't be assumed that the explanation of the success of the renormalization transformation in QFT is exactly the same in all respects as the explanation for its success in the condensed matter case, and thus adopting a reductionist explanation of the transformation in condensed matter does not necessarily compel us to accept a reductionist approach in QFT as well.

\section{Ontologies \label{ontologies}}

Suppose we agree at least  in principle that there are reasons to think the supervenience of the large on the small may fail in the regimes relevant to QFT. Still, in order for this idea to gain any traction it will be necessary to demonstrate that there is some possible way to implement such a thing within a physically plausible model. 

  One possibility would be to   simply advocate something akin to the tower of effective field theories, in which we say that  all the quantum field theories defined at different scales are equally fundamental. In this picture, all of the QFTs at different scales exist simultaneously, as real elements of the physical world, but none `gives rise' to others. An approach of this kind is advocated by \cite{cd3bcfc7-c1f3-3ac4-a548-136b57165545}. Alternatively, one could also imagine an approach in which we deny the existence of meaningful links between scales entirely. This would involve a Cartwright-style `patchwork,' in which scientific laws apply only within their own limited domain of validity and science as a whole is disunified \citep{Cartwright}.

However, a number of the points  covered in section \ref{lats} appear to give reason not only to deny the supervenience of the large on the small in QFT, but to actually \emph{reverse} the direction of fundamentality! For example, the discussion of naturalness suggested a solution which involves not only denying that  $\Lambda_0$ and $c$ are fundamental, but also stipulating that the value $\Lambda_M$ defined at larger distances \emph{is} fundamental. So in this section I will seek to understand what it might look like to implement a reversal of the direction of fundamentality in a physically plausible realist model.  

There are probably many ways this could be done, but one natural possibility is to make use of an existing interpretation of non-relativistic quantum mechanics. In particular, there are a number of `$\psi$-nonphysical' interpretations of quantum mechanics which suggest that `the quantum state' is not an objective element of physical reality - it is merely a useful tool to predict the behaviour of the actual elements of physical reality\footnote{Sometimes approaches of this kind are known as `psi-epistemic.' However, I find this term problematic since the quantum state may be non-physical without necessarily being just a description of somebody's knowledge.}. There is work to be done in establishing exactly how these approaches can be extended to quantum field theory, but it seems natural to suppose that in such a view the quantum fields, or at least the quantum state of the fields, would not be understood as  elements of physical reality, so indeed we would potentially not need the very small-distance `bare theory' to play a literal representative role.  So here  I will explore what a non-reductionist approach might look like in the context of two different $\psi$-nonphysical approaches.

\subsection{Relational Approaches} 

Relational approaches to quantum theory, most famously Rovelli's Relational Quantum Mechanics (RQM), have been around for some time and have a number of appealing features \citep{1996cr}. These approaches have mostly been applied to non-relativistic   quantum mechanics, so some work must still be done to show explicitly how to extend them to QFT, but in this section I will make no attempt to give a fully fleshed out relational account of QFT - instead I will imagine a somewhat generic relational approach and consider how it would deal with the issues considered  in this paper. 

So in the relational spirit, let us  say that the quantum state of a system or field is relativized to an observer, and a single system or field may have a different state relative to different observers. And let us follow Rovelli in using the term `observer' in a very general way, such that in principle any physical system can play the role of an observer. We can then extrapolate beyond standard relational approaches by saying that since dynamics describes how states change over time, it is natural in the relational picture to say that a system or field may have different \emph{dynamics} relative to different observers. And therefore we may suggest that since the dynamics of a quantum field theory typically depend on the scale at which we probe the theory, it is natural in a relational approach to QFT to interpret the relevant scale as the scale \emph{associated with the observer to which the dynamics are relativized}. So for example, if system $S$ has energy $E$, then when it interacts with a quantum field, it will  `see' the quantum field  at energy $E$ (which, recall, corresponds to some particular distance scale $D$) - the state of the field relative to $S$ will be the state defined at the distance scale $D$, and that state will undergo evolution according to the Lagrangian defined at the distance scale $D$, with the appropriate masses and coupling constants for distance scale $D$.   

\cite{pittphilsci18056} makes a somewhat similar proposal:  `\emph{I will conclude briefly with a more radical suggestion to circumvent this issue: namely, to modify the standard semantic tenet of scientific realism endorsed by selective realists (e.g., Psillos 1999; Chakravartty 2007) and index (approximate) truth to physical scales.}' Rivat's concern here is to articulate a form of realism about EFTs which allows the realist to be committed to something relatively concrete, so he recommends truth indexed to a scale in order that the realist can commit to continuum fields, even though we know that  EFTs most likely cease to be accurate as one takes the continuum limit - the thought is that a continuum field can be a literally true description of goings-on within some range of scales, without any need to suppose that the field has properties on arbitrarily short scales or indeed anywhere outside the relevant range of scales.   The relational approach suggested above achieves something similar, but in an arguably less radical way - we are leaving the notion of truth alone and simply arguing that what is true about the world is that physical descriptions are always relativized to an `observer' and thus are necessarily indexed to scales. 
 
The relational approach has  several nice consequences. First, there are no physical systems with infinite energy, so no system can possibly probe a quantum field at infinitely small scales, as this would require infinite energy. It  follows that in this relational picture, the continuum theory associated with any QFT is physically meaningless: no system will ever evolve relative to another system according to the continuum theory defined at unreachably high energies, so infinities appearing in the continuum theory do not describe anything in reality.   And second, the relational picture marries particularly well with Butterfield's observation that `\emph{we can't take the system out of the field of force}.' Indeed, the proponent of a relational picture may be tempted to make a stronger statement: some proponents of relational views hold that systems do not have intrinsic properties in and of themselves, they only have relational properties, so counterfactuals about what we would observe if we could take the system out of the field are simply nonsensical, since the system cannot be expected to have individual, non-relative properties to define those counterfactuals.

\subsection{The Final-Measurement Approach}

Another $\psi$-nonphysical approach which might be useful in this context is Kent's `Lorentzian solution to the quantum reality problem.' In Kent's picture,  there is no collapse of the wavefunction; we just allow the wavefunction to undergo its standard unitary evolution over the whole course of history, and then, at the end of time, we imagine a single measurement being performed on the final state to determine the actual content of reality:  in Kent's words, `\emph{an event occurs if and only if it leaves effective records in the final time ... measurement}' \cite{Kent}.  Note that the final measurement need not be understood literally, as it simply describes the process in which a course of history is selected and actualised. For example,    \cite{  2017Kent} proposes a final measurement on the positions of photons which have been reflected off matter at various points, and then  defines the actual content of reality in terms of `beables' defined  at an individual spacetime point  $x$ by the expectation value of some operators (e.g. the stress-energy tensor components) at  $x$ conditional  on the detections of photos outside the future lightcone of $x$.  

 As discussed by Kent in  \cite{Kent,2015KentL,2017Kent}, decoherence plays an important role in this picture, because decoherence ensures that macroscopic events are recorded robustly in their environment, so they will almost always leave effective records in the final time measurement; thus for any macroscopic event the procedure for defining beables will typically represent the systems involved in that event as being in definite states. Whereas for microscopic quantum events which are appropriately shielded from decoherence, there will not usually be robust environmental records and thus interference effects can subsequently erase all traces of such events, so the resulting beables will not represent the systems involved as being in definite states. Thus the final-measurement approach predicts that macroscopic happenings occur in a single definite way whereas most microscopic happenings occur in an indefinite way or do not really occur at all. That is, the mechanism by which the beables are selected ensures that in general, beables will be associated with events at small distances only when they are magnified to larger scales by some kind of `probe' system like a measuring instrument, since this ensures that there will be robust ongoing records which have a good chance of surviving until the final measurement. 

Kent's model is non-reductionist in an interesting way. The model can still in some sense be defined in terms that seem consistent with petty reductionism, since the beables are well-defined at each spacetime point and larger structures are simply constructed out of the distribution of   beables at individual spacetime points. But the model does have  the consequence  that in general,  microscopic phenomena  exist only when some process magnifies microscopic effects up to macroscopic scales. So for example, small-scale phenomena studied in particle physics will not be taking place all the time in this model; they are only  brought into being during scattering experiments at high energies, or other processes which magnify microscopic effects up to macroscopic effects. Thus there is a non-reductionist flavour to the model - in particular, it seems natural to say that in such a model it  is not really the case that all  matter is really made up of fundamental quantum fields or quarks or something of that nature, since   the quantum fields and quarks are \emph{created} only in special circumstances when they become relevant to macroscopic measurement outcomes or other macroscopic effects. 

This non-reductionist facet of Kent's model allows it to solve the problem of  infinities in a similar way to the relational approach. For it tells us that  small-distance phenomena   are actualized only in the case where the effects are magnified up to larger scales by appropriate probes; and since there are no doubt robust physical limitations on the scales to which a realistic instrument can probe, we will never have to apply the theory at the smallest possible distances. Indeed, there is something like a `cutoff' in this picture, but it is not associated with new physics - it is merely the point at which the existing physics becomes meaningless since physics below this scale not accessible to any mechanism which could actualize the beables. This `cutoff' need not occur at any specific physical scale - all that is necessary to solve the infinity problem is that although we may asymptote towards smaller and smaller scales, we will never actually reach the infinities.

\section{Conclusion}

In this article I have argued, first,   that there are some indications that ontological reductionism is not the right paradigm in which to understand  quantum field theory, and second, that the structure of QFT does not compel us to accept the supervenience of the large on the small. There are a variety of conceptual problems connected with the reductionist formulation of QFT which seem like they could possibly be relieved by reinterpreting the formalism in a way which doesn't assume the smallest  distance scales are always the most fundamental - for example, this could  help address long-standing worries about infinities and naturalness. Certainly it seems hard to deny that something unusual is going on with the methodology of reductionism in the context of QFT, which suggests that this would be a sensible point at which  to re-evaluate our commitment to ontological reductionism. 
 
Of course care must be taken with this kind of project. One natural concern would involve the possibility that what is going on here is just a confusion  between the direction of epistemic and ontic dependence. After all, when we do science we are inevitably making observations about larger scales and using them to infer about what is going on at smaller scales, so our theoretical descriptions of small-scale effects will always have some degree of dependence on what is going on at larger scales, because it is only via large-scale effects that we can learn about and functionally define the  smaller-scale entities. In light of this, any proposal that nature may exhibit some non-reductionist features will potentially be subject to the criticism that these features are just artifacts of our epistemic limitations and not real indications that reductionism is breaking down. This is a very reasonable concern, and I will not try to refute it here; for my goal is not to show that ontological reductionism definitely fails in QFT, but simply to make the point that our current understanding of QFT suggests that this possibility merits investigation.

One might also have worries about the unity of science: for those realists who find a Cartwright-style `patchwork' view of science unpalatable, it may seem impossible to accept that macroscopic things can supervene on microscopic things but also very small sub-microscopic things supervene on somewhat larger microscopic or sub-microscopic things, as this would seem to involve some kind of sudden, discontinuous change in the direction of ontological dependence which prima facie looks incompatible with a commitment to a single, unified reality. However, while models of this kind wouldn't necessarily be unified in the same way as traditional models in the reductionist paradigm, they need not be unacceptably disunified. Reduction to the smallest of scales is not necessarily the only way in which reality could be unified, and regardless of the outcome, it could be interesting to explore alternative paradigms.

\end{document}